\def\draftversion{false}

\RequirePackage{ifthen}
\ifthenelse{\equal{\draftversion}{true}}{
	\documentclass[aps,prl,10pt,galley,amsmath,amssymb,
	floatfix,
	citeautoscript]{revtex4}
}{
	\documentclass[aps,prl,10pt,twocolumn,amsmath,amssymb,
	citeautoscript]{revtex4-1}
}

\usepackage{graphicx}
\usepackage[usenames,dvipsnames]{color} 
\usepackage{bm} 
\usepackage[update,prepend]{epstopdf}
\usepackage{soul} 
\usepackage{dcolumn} 
\usepackage{comment}
\usepackage{gensymb}
\usepackage[linktocpage=true]{hyperref}
\usepackage{hyperref}
\usepackage{comment}
\usepackage{notes2bib}
\hypersetup{
  pdfnewwindow=true, colorlinks=true,
  linkcolor=blue, anchorcolor=blue,
  citecolor=blue, filecolor=blue,
  menucolor=blue, urlcolor=blue}

\def\comment#1{}

\ifthenelse{\equal{\draftversion}{true}}{
	\marginparwidth 2.7in
	\marginparsep 0.5in
	\newcounter{comm} 
	\def\commnext{\stepcounter{comm}}
	\def\commtext{{\bf\color{blue}[\arabic{comm}]}}
	\def\commmar{{\bf\color{blue}[\arabic{comm}]}}
	\def\dvm#1{\commnext\marginpar{\small DV\commmar: #1}\commtext}
	\def\jkm#1{\commnext\marginpar{\small JK\commmar: #1}\commtext}
	\def\krm#1{\commnext\marginpar{\small KR\commmar: #1}\commtext}

}{
	\def\jkm#1{}
	\def\krm#1{}
    \def\dvm#1{}
	
}

\begin{document}


\title{Engineering Weyl phases and nonlinear Hall effects in T$_d$-MoTe$_2$}

\author{Sobhit Singh}
\email{sobhit.singh@rutgers.edu}
\author{Jinwoong Kim}
\author{Karin M. Rabe}
\author{David Vanderbilt}
\affiliation{
  Department of Physics and Astronomy,
  Rutgers University,
  Piscataway, New Jersey 08854-8019, USA
}


\begin{abstract}
MoTe$_2$ has recently attracted much attention due to the observation of
pressure-induced superconductivity, exotic topological phase transitions, and
nonlinear quantum effects. However, there has been
debate on the intriguing structural phase transitions among various observed phases
of MoTe$_2$, and their connection to the underlying topological electronic properties.
In this work, by means of density-functional theory (DFT+U) calculations,
we investigate the structural phase transition between the
polar T$_d$ and nonpolar 1T$'$ phases of MoTe$_2$ in reference
to a hypothetical high-symmetry T$_0$ phase
that exhibits higher-order topological features.
In the T$_d$ phase we obtain a total of 12 Weyl points,
which can be created/annihilated, dynamically manipulated, and switched
by tuning a polar phonon mode. We also report the
existence of a tunable nonlinear Hall effect in T$_d$-MoTe$_2$, and propose 
the use of this effect as a probe
for the detection of polarity orientation in polar (semi)metals.
By studying the role of dimensionality, we identify a configuration in which a
nonlinear surface response current emerges.
The potential technological applications of the tunable Weyl phase
and the nonlinear Hall effect are discussed.
\end{abstract}

\maketitle

Owing to its intriguing structural and electronic phase transitions
and novel technological applications, MoTe$_2$ remains in the active
area of research~\cite{QingWang_NatRev2012, KeumNatPhy2015, KenanZhangNatComm2016, YWangNature2017_electrostaticdoping, Wang_RevModPhys2018, ABerger_npj2018, JWencanPRB2018, SiChenNanoLett2019}.
Specifically, the experimentally tunable
structural phase transition between the 1T$'$ and
T$_d$-phases~\cite{CHeikes_PRM2018, ShuoguoY_NatComm2019, huang2019polar}
allows for the exploitation of the topological electronic properties
and the electronic phase transitions, yielding various novel
quantum phenomena such as extremely
large magnetoresistance~\cite{FCChenPRB2016,QLPei_PRB2017,
Thirupathaiah_PRB2017, SangyunLee_SciRep2018},
various kinds of Hall effects~\cite{FCChenPRB2018,YZhang_2DMat_2018,Qian1344,SejoonLimPRB2018,Binghai2018,JZhou_PRB2019},
complex Fermiology~\cite{DRhodes_PRB2017, AWeber_PRL2018, Aryal_PRB2019},
and tunable polar/phase
domain walls in MoTe$_2$~\cite{huang2019polar}.
Recent works have confirmed the existence of
Weyl fermions in T$_d$-MoTe$_2$, and reported topological
quantum oscillations and pressure-enhanced
superconductivity~\cite{YanSun_PRB2015, Soluyanov_Nat2015, LunanHuang_NatMat2016,
Zhijun_PRL2016, KeDengNatPhys2016, ATamai_PRX2016, DRhodes_PRB2017, CrepaldiPRB2017,
Takahashi_PRB2017, Yanpeng_NatComm2016, CHeikes_PRM2018, ABerger_npj2018, ZAnminPRB2019}.
However, the total number and location of Weyl fermions in MoTe$_2$ are
still under debate~\cite{YanSun_PRB2015, Zhijun_PRL2016, DRhodes_PRB2017, AWeber_PRL2018, CrepaldiPRB2017, NXu_PRL2018, Aryal_PRB2019, AWeber_PRL2018}.
Strikingly, a higher-order topological phase has been predicted in
1T$'$-MoTe$_2$~\cite{wang2019higher, FengTang_NatPhy2019, MotohikoE_SciRep2019}.

There has recently been substantial progress in understanding the 
topological electronic properties of MoTe$_2$~\cite{YanSun_PRB2015, Soluyanov_Nat2015, LunanHuang_NatMat2016, Zhijun_PRL2016, ATamai_PRX2016, DRhodes_PRB2017, CrepaldiPRB2017, Takahashi_PRB2017, Yanpeng_NatComm2016, CHeikes_PRM2018, NXu_PRL2018, ZAnminPRB2019, wang2019higher}. 
Most of the reported theoretical and experimental studies have focused
on the characterization of the distinct phases of MoTe$_2$~\cite{SYChen_NL2016, HJKimPRB2017, MYZhang_PRX2019, CHeikes_PRM2018, MYZhang_PRX2019, RuiHePRB2018, PTsipasAFM2018, YuTaoPRB2019, DissanayakeNPJ2019}.
In contrast, the link between the higher-order topological phase
(1T$'$) and the Weyl phase (T$_d$) has been relatively unexplored. Specifically, a systematic connection among these electronic phases, in the context of the potential energy surface profile and crystal symmetries, has not been clarified in the literature.

In this work, by means of {\it ab-initio} density-functional theory (DFT)
calculations, we first investigate the structural phase transition
between the polar T$_d$ and nonpolar 1T$'$ phases of MoTe$_2$ using a
hypothetical reference phase T$_{0}$ introduced by us in the context of
the experimental results in Ref.~\cite{huang2019polar}. 
In particular, we focus on the electronic phase transitions occurring in the vicinity
of the T$_{0}$ phase. 
We study the evolution of the Weyl points (WPs) in the polar phase along the
polarity reversal path, and demonstrate that a higher-order topological phase
naturally appears when all WPs annihilate each other at zero polar distortion in the reference phase T$_{0}$.
We also report on the existence of a tunable nonlinear
Hall effect and propose that the nonlinear Hall effect can be used to detect polarization direction and switching in polar metals or semimetals,
especially those with strong sources of Berry curvature near the Fermi energy. 
Such a tunable nonlinear Hall effect could lead to electrically
switchable circular photogalvanic~\cite{SYXu_NatPhy2018, SejoonLimPRB2018},
bulk rectification~\cite{Ideue_NatPhy2017}, and chiral polaritonic effects~\cite{Basovaag_Sc2016}.
Finally, we discuss the role of dimensionality on this effect.
We notice that the surface termination along the (001) direction
leads to the manifestation of a nonlinear surface response
current solely arising due to the broken symmetries at the surface.

\begin{figure}[htb]
 \centering
 \includegraphics[width= 8cm]{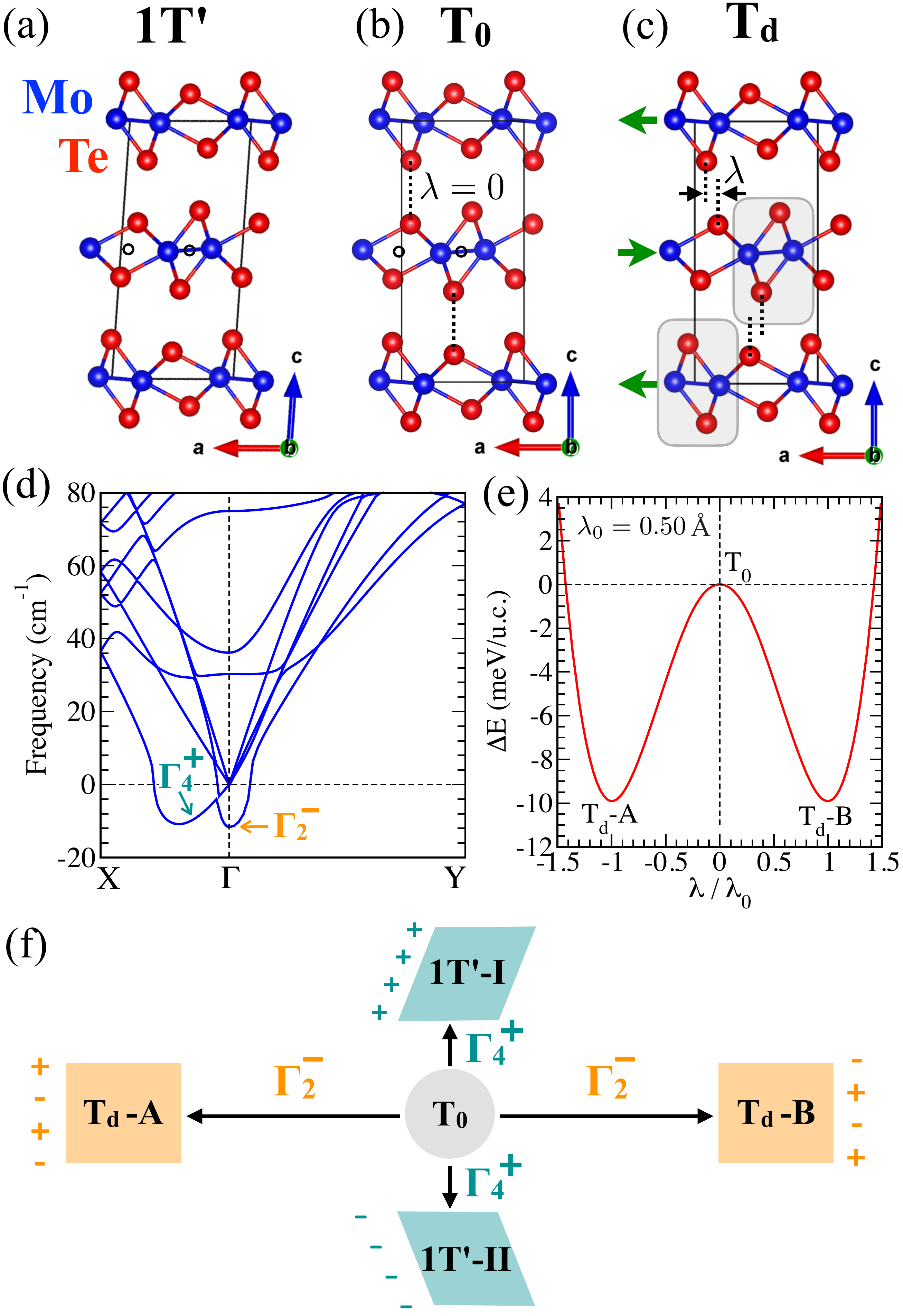}
 \caption{Crystal structure of MoTe$_2$ in (a) 1T$'$,
 (b) designed T$_{0}$, and (c) T$_{d}$ phases.
 Green arrows denote the interlayer displacement direction
 parallel ($+$) or antiparallel ($-$) to $\vec{a}$, and
  $\lambda$ represents the interlayer displacement parameter (see text).
 Hollow circles `$\circ$' mark the inversion centers in the 1T$'$ and T$_0$ phases.
 (d) The phonon band structure of T$_{0}$ phase
  shown within [-20, 80]\,$\rm{cm^{-1}}$.
 (e) The double-well potential energy profile of
 T$_{0}$ phase as a function of the inversion
 symmetry breaking parameter $\lambda$.
 (f) A schematic representation showing the link
 between all T$_d$ and 1T$'$ phases {\it via}
 the reference structure T$_{0}$.
A 2D map of the potential energy surface as a function of  $\Gamma_{2}^{-}$ and $\Gamma_{4}^{+}$ modes is given in Ref.~\cite{huang2019polar}. 
}
 \label{fig:struct}
 \end{figure}

MoTe$_2$ crystallizes in three distinct phases:
(i) 2H (hexagonal, $P6_3/mmc$),
(ii) 1T$'$ (monoclinic, $P2_1/m$), and
(iii) T$_d$ (orthorhombic, $Pnm2_1$)~\cite{RClarke1978,
HJKimPRB2017, CHeikes_PRM2018}.
In all three phases, Mo and Te atoms form Te-Mo-Te triple layers,
which stack along the $c$-axis and interact {\it via} weak van der Waals interactions.
The Te atoms form symmetrical polyhedra in the hexagonal 2H phase,
whereas these polyhedra are markedly distorted in the
1T$'$ and T$_d$ phases~\cite{huang2019polar}, as shown in Fig.~\ref{fig:struct}(a,c).
Both phases are quite similar, except for the
fact that 1T$'$ is monoclinic ($\beta \not=$ 90) while
T$_d$ is orthorhombic ($\alpha = \beta = \gamma = 90$).
In both phases, Mo atoms dimerize forming long-short
bonds along the $\vec{a}$ lattice vector and zigzag
Mo-Mo metallic bonds running along the $\vec{b}$ direction.

We notice an interesting symmetry between the Mo-Te polyhedra
(see light grey rectangles in Fig.~\ref{fig:struct}(c))
of alternating triple layers.
These polyhedra alternatively adopt either clockwise or
counterclockwise twist (as viewed along $\vec{b}$)
in the alternating triple layers.
In the T$_d$ phase, adjacent layers are
connected by $\mathcal{M}_x\,| \{\mathcal{T}({\frac{\vec{a}}{2}}(1 + \lambda))\}$
symmetry operation, where $\mathcal{M}_x$ is a vertical mirror,
$\mathcal{T}({\vec{a}/{2}})$ denotes translation by $\vec{a}/2$,
and $\lambda$ denotes an interlayer displacement along $\vec{a}$,
as shown in Fig.~\ref{fig:struct}(c).
The main cause of the nonzero $\lambda$ is the presence of
steric interactions between Te atoms in the adjacent triple layers,
which drive an in-plane shift of the alternating
layers along $\vec{a}$ so as to increase the separation between
these atoms.
Taking the above facts into account, we define a
nonpolar high-symmetry phase T$_{0}$ ($Pnma$)
having $\lambda$ = 0, as shown in Fig.~\ref{fig:struct}(b).

Fig.~\ref{fig:struct}(d) shows an enlarged phonon spectrum of the T$_{0}$ phase.
The full phonon spectra of T$_{0}$, 1T$'$, and T$_d$ phases together
with all the theoretical details are given in the supplemental
material (SM)~\cite{SM}. We notice only two phonon instabilities in the T$_{0}$ phase:
(i) an unstable optical zone center phonon mode
($\Gamma_{2}^{-}$), and (ii) a linearly-dispersing unstable
phonon branch along $\Gamma$-X direction indicating an
elastic instability ($\Gamma_{4}^{+}$).
The first instability, $\Gamma_{2}^{-}$, breaks the inversion
symmetry of the T$_{0}$ phase and corresponds to an in-plane optical
vibration of the alternating triple layers.
By modulating T$_{0}$ phase along $\Gamma_{2}^{-}$ mode,
we obtain a double-well potential
energy profile with two local minima at
$\lambda = \pm 0.50$\,\AA, as shown in Fig.~\ref{fig:struct}(e).
These local minima belong to the two polar variants of the T$_d$
phase, which we refer as T$_d$-A and T$_d$-B.
The interlayer displacement pattern of the alternating Mo-Te triple layers
in the T$_d$-A and T$_d$-B phases
is $-+-+...$ and $+-+-...$, respectively, thus,
ensuring the orthogonality of the T$_d$ phase.
On the other hand, the elastic instability ($\Gamma_{4}^{+}$ mode)
causes a shear distortion of the unit cell, resulting in two ferroelastic
twin phases, 1T$'$-I and 1T$'$-II, corresponding to the interlayer displacement pattern of
 $\rm{++++...}$ and $\rm{----...}$, respectively (details in SM~\cite{SM}).
Fig.~\ref{fig:struct}(f) schematically represents
the connection between two polar T$_d$ phases and two
ferroelastic 1T$'$ phases in reference to the unstable high symmetry T$_{0}$ phase.

\begin{figure}[htb]
 \centering
 \includegraphics[width= 6.6cm]{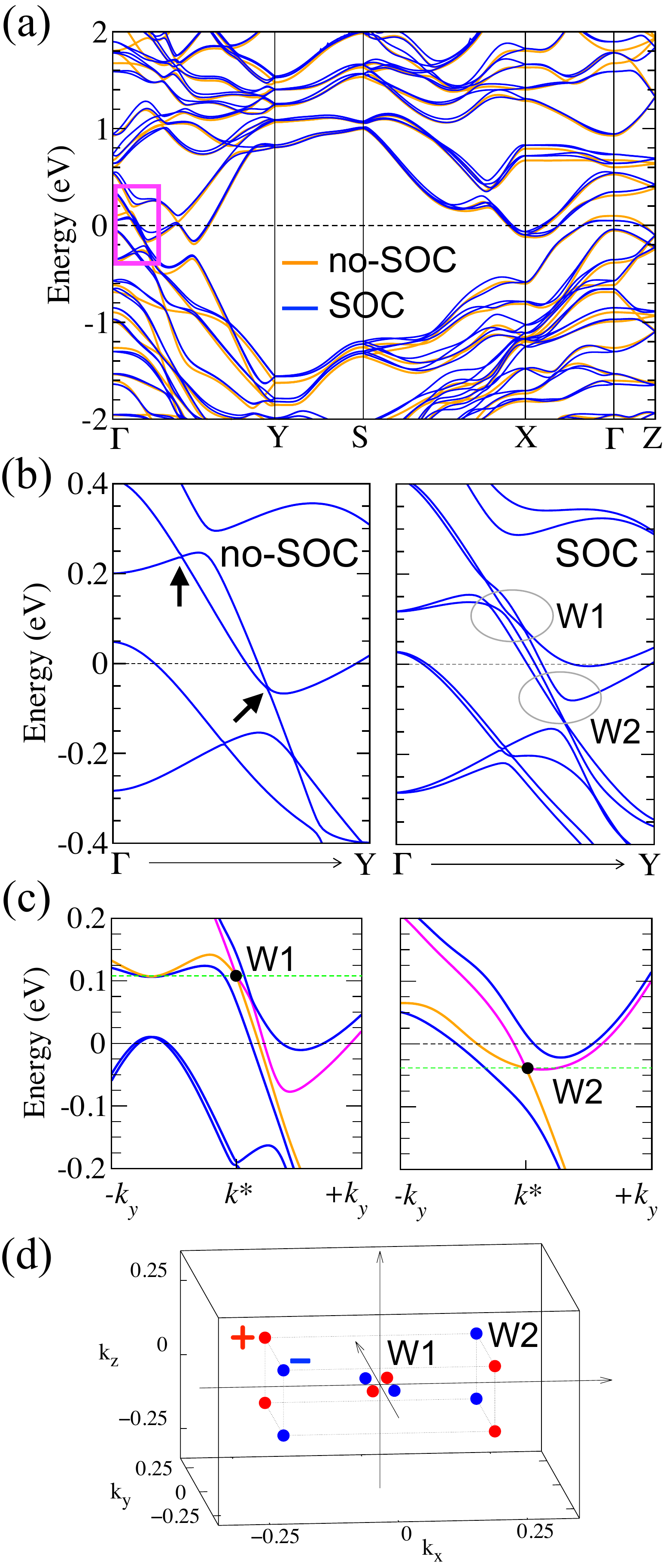}
 \caption{(a) Electronic band structure of T$_d$-MoTe$_2$
 calculated with (blue) and without (orange) SOC.
 (b) Enlarged view of bands calculated
 near the Fermi level $E_F$ along the $\Gamma$-Y path highlighted 
 by the magenta rectangle in (a). Grey ovals in the right panel of (b) show
 the regions near which Weyl crossings occur.
(c) Band dispersions (with SOC) along a momentum cut parallel to $k_y$ passing through the Weyl point (black dots) in question (see SM~\cite{SM}); both W1 and W2 are of type II.
The dotted black horizontal line represents the Fermi level, and 
the green dotted line marks the energy of WPs situated at $k^{*}$.
(d) Distribution of all WPs in the BZ for the T$_d$-A phase;
 red/blue dots depict $\rm{+/-}$ chiralities of the WPs. } 
 \label{fig:bands}
 \end{figure}

Due to the broken inversion symmetry requirement,
the path connecting the 1T'-I and 1T'-II phases cannot access the Weyl phase.
Therefore, we focus on the T$_d$-A $\rightarrow$ T$_d$-B path,
investigating the subtle changes in the electronic band
structure that occur there.
Without spin-orbit coupling (SOC),
the lowest conduction band and the highest valence band
cross each other near the Fermi level,
forming gapless nodal loops~\cite{Zhijun_PRL2016}
 above and below $E_{F}$, as marked by black arrows in Fig.~\ref{fig:bands}(b)
 ~\bibnote{The states near the Fermi energy ($E_{F}$) are mainly
composed of Mo-$4d$ and Te-$5p$ orbitals. Due
to the semicorrelated nature of Mo-$4d$ orbitals, pure DFT fails
to correctly describe the Angle-Resolved Photoemission
Spectroscopy (ARPES) data and pressure dependence of
quantum oscillation frequency measurements in
MoTe$_2$~\cite{NXu_PRL2018, Aryal_PRB2019, SKimuraPRB2019}.
Adding an on-site Hubbard term $({U}_{\mathrm{eff}})$,
the Hubbard term, on Mo-$4d$ orbitals has been reported
to solve this issue. Therefore, we consider
${U}_{\mathrm{eff}}$ = 2.4\,eV as suggested by Xu et al.~\cite{NXu_PRL2018}}.
Inclusion of SOC destroys the nodal loops and results in discrete
WPs formed away from the high-symmetry directions near the Fermi level.
We find that there are two sets of WPs: 
(i) W1 WPs lying above the Fermi level at $E_{F}+$0.108\,eV (in the $k_{z}=0$ plane), and (ii) W2 WPs lying below the Fermi level at $E_{F}-$0.038\,eV (off the $k_z\!=\!0$ plane).
A detailed examination of the band dispersion in the vicinity of
the WPs, as shown in Fig.~\ref{fig:bands}(c), reveals that 
both W1 and W2 WPs have type-II nature. 
W1 WPs have a stronger tilting of Weyl cone compared to W2.  
We note that a subtle lattice distortion may change the type of WPs, as suggested by Sun et al.~\cite{YanSun_PRB2015}. 
A careful investigation of the electronic band structure
reveals a total of 12 WPs (4 W1 and 8 W2) in the full Brillouin zone (BZ),
as shown in Fig.~\ref{fig:bands}(d).
Coordinates of all WPs are provided in the SM~\cite{SM}.
The T$_d$-A and T$_d$-B phases host exactly the same number of WPs at the same
coordinates in momentum and energy, but with reversed chirality.

Motivated by the above results, we investigate the
evolution of the WPs along the T$_d$-A $\rightarrow$ T$_0$ $\rightarrow$ T$_d$-B
path as a function of $\lambda$.
We observe that WPs get created in pairs
as we move away from the T$_d$-A phase.
The total number of WPs increases from 12 to 16 and then 20, 24, 28, and 32, as
we vary $|\lambda/\lambda_{0}|$ from 1.0 to 0.92, 0.88, 0.79, 0.72, and 0.63,
respectively ($\lambda_{0}$ = 0.50\,\AA)~\cite{SM}.
The maximum number of obtained WPs is 32.
This finding explains why previous authors reported such different counts of
the number of Weyl points~\cite{YanSun_PRB2015, Zhijun_PRL2016, DRhodes_PRB2017, AWeber_PRL2018, CrepaldiPRB2017, NXu_PRL2018, Aryal_PRB2019, AWeber_PRL2018}, and reveals that the total number of
WPs in MoTe$_2$ is very sensitive to the lattice distortions.
As we further tune $|\lambda/\lambda_{0}|$, the WPs move towards
their opposite partners in momentum space and start pair-annihilating,
leaving no remaining WPs at $|\lambda/\lambda_{0}|$  = 0 (at T$_{0}$ phase).

Due to the absence of WPs and the presence of a double band inversion
at the $\Gamma$ point, the T$_{0}$ phase turns into a
second-order topological insulator, similar to the
1T$'$ phase~\cite{FengTang_NatPhy2019, wang2019higher}. Notably,
we find that both the T$_{0}$ and 1T$'$ phases belong
to a strong topological class 20 as classified in Ref.~[\onlinecite{vasp2trace2019}]
having topological invariant $z_{4}$=2.
As we cross the T$_0$ phase and migrate towards the T$_d$-B phase,
the WPs systematically start reappearing, and the aforementioned process
repeats but with the switched chirality of
WPs~\bibnote{An animation showing the evolution of WPs as a function of $|\lambda/\lambda_{0}|$ is provided in the SM~\cite{SM}.}.
The pairwise creation/annihilation of WPs causes
abrupt changes in the Berry curvature and Fermi-surface
geometry yielding a nonzero Berry curvature
dipole moment (BCDM)~\cite{Sodemann} and, as a result, a nonlinear Hall effect in
T$_d$-MoTe$_2$~\cite{Binghai2018}, as we discuss below.

In the study of the nonlinear quantum Hall effect~\cite{Sodemann, ZZDuPRL2018, ZZDu_Natcomm2019, MatsyshynPRL2019},
a transverse current is predicted to be generated by a harmonically
oscillating electric field $E_c = \textrm{Re} \{ \mathcal{E}_c
e^{i\omega t} \}$ in the absence of inversion symmetry.
The response current up to second order reads
$j_a = \textrm{Re} \{ j_a^{0} + j_a^{2\omega} e^{2i\omega t} \} $,
where a rectified current
$j_a^{0} = \chi_{abc}\, \mathcal{E}_b \mathcal{E}_{c}^{*}$ and
a second-harmonic
current $j_a^{2\omega} = \chi_{abc}\, \mathcal{E}_b \mathcal{E}_c$ depend on
the nonlinear conductivity tensor $\chi_{abc}$, where
$a,b,c \in \{x,y,z\}$.
The nonlinear conductivity tensor associated with the
BCDM ($D_{bd}$) can be written as
\begin{eqnarray}
\chi _{abc} (\omega)&=& -\varepsilon _{adc} \frac{e^3 \tau}{2(1+i\omega \tau)}D_{bd},
\label{eq:chi}
\end{eqnarray}
where $\varepsilon _{abc}$ is the rank-three Levi-Civita symbol
and $\tau$ is the relaxation time.
The $D_{bd}$ is obtained by integrating
Berry curvature, weighted by the Cartesian component
of the group velocity on the Fermi surface according to
\begin{eqnarray}
D_{bd} &=& \oint_{\rm FS} \frac{d^2 \bm{k}}{(2\pi)^3} \sum_{n}
v^{n}_{b}(\bm{k})
\Omega_d^n(\bm{k}),
\label{eq:bcdm}
\end{eqnarray}
where $v^{n}_{b}(\bm{k})=\partial_{k_b} E_{n\bm{k}}/{|\nabla_{\bm k} E_{n\bm{k}}|}$
is a normalized group velocity component for band $n$,
and $\mathbf{\Omega}^n$ is the Berry curvature pseudovector
defined {\it via} $\Omega_{bc}^n=\varepsilon _{abc}\,\Omega_a^n$.
The superscripts represent band indices.
We compute the Berry curvature using the Kubo formula
\begin{equation}
\Omega _{ab}^{n}(\bm{k}) = -2\hbar^2\sum_{m \neq n} \textrm{Im}
\frac{\left<n\bm{k} | \hat{v}_a | m\bm{k}\right>
	\left<m\bm{k} | \hat{v}_b | n\bm{k}\right>}
{\left( E_{n\bm{k}} - E_{m\bm{k}} \right) ^{2} + \delta ^2 },
\end{equation}
where $\hat{v}_a$ is the velocity operator and
$\delta\!=\!0.1$\,meV is a broadening term (see~\bibnote{Due to the heavy
  computational cost of the Kubo formula and slow convergence of
  BCDM with respect to the $k$-mesh size, we first compute the Fermi
  surface by employing the tetrahedron method at a given $k$-grid,
  and sample Berry curvature only at the reduced grid points near
  the Fermi surface.  The convergence of BCDM was achieved at a
  $k$-grid of size $278 \times 510 \times 130$ with Gaussian smearing,
  where the broadening width corresponds to $\sim$50\,K.}
for numerical details).

In the presence of inversion symmetry, i.e.,
the case of 1T$'$-MoTe$_2$, the BCDM
completely vanishes. Instead, in the polar
T$_d$ phase, a non-vanishing
BCDM is allowed~\cite{Sodemann, QiongNature2019}.
T$_d$-MoTe$_2$ exhibits simple mirror
$\mathcal{M}_y$ and glide mirror
$\mathcal{M}_{x}\mathcal{T}({\vec{c}/2})$ symmetries,
exerting constraints on the BCDM
tensor. For instance, $\mathcal{M}_y$, a mirror plane
normal to the chain direction, forces
the group velocity $v_a$ and Berry curvature
$\Omega_b$ to obey
\begin{eqnarray}
\mathcal{M}_{y}:\, (v_x, v_y, v_z) &\rightarrow& (v_x, -v_y, v_z)\\
(\Omega_x, \Omega_y, \Omega_z) &\rightarrow&
(-\Omega_x, \Omega_y, -\Omega_z).
\end{eqnarray}
Here, the $v_i$ denotes group velocity for a particular band
at a particular $\bm k$.
Thus, under the $\mathcal{M}_y$ symmetry operation,
all components of the BCDM
tensor vanish except for the $D_{xy}$, $D_{yx}$, $D_{yz}$, and $D_{zy}$
terms. A further consideration of
$\mathcal{M}_{x}\mathcal{T}({\vec{c}/2)}$ symmetry eliminates the
$D_{yz}$ and $D_{zy}$ terms as well. Thus, only two
terms, $D_{xy}$ and $D_{yx}$, survive in T$_d$-MoTe$_2$.

\begin{figure}[htb]
 \centering
 \includegraphics[width= 7.6cm]{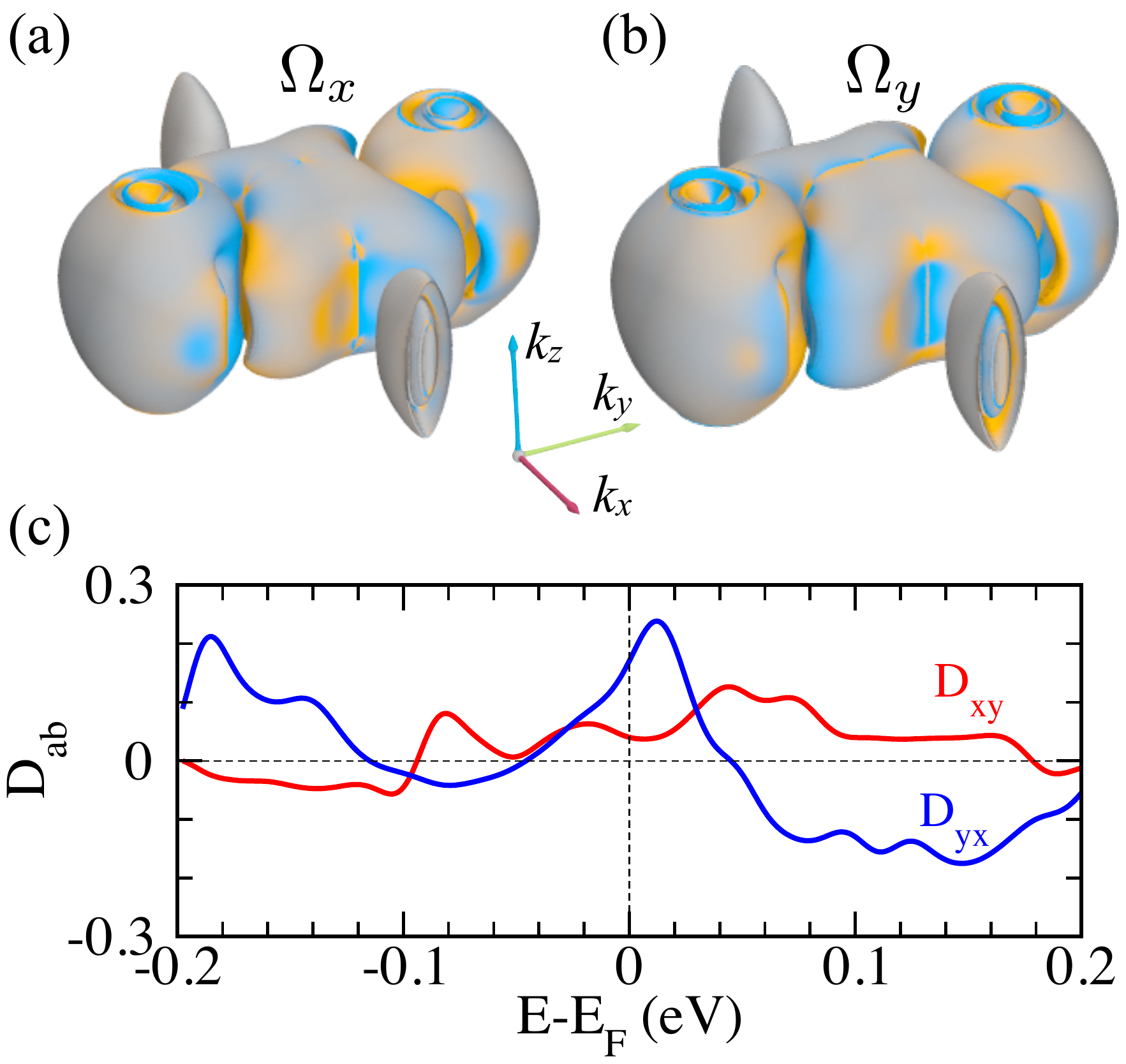}
 \caption{ Calculated Berry curvature (a) $\Omega_{x}$ and
        (b) $\Omega_{y}$ on the Fermi surface
		of MoTe$_2$ in T$_d$-A phase. Yellow (Blue) color
		represents positive (negative) Berry curvature.
		(c) Calculated BCDM of
		MoTe$_2$ in T$_d$-A phase. The non-vanishing
		$D_{xy}$ and $D_{yz}$ terms are plotted with
		respect to the chemical potential. }
 \label{fig:BCDM}
 \end{figure}

The nonvanishing nature of the $D_{xy}$ and $D_{yx}$ terms
can be anticipated from the Berry curvature distribution plot on the Fermi surface [Fig.~\ref{fig:BCDM}(a-b)].
Because of the complex metallic bands with anisotropic
group velocities in type-II Weyl semimetals,
the Fermi surface has significant Berry curvature
even away from the WPs (see SM~\cite{SM}). This renders the BCDM
more sensitive to the chemical potential than for type-I Weyl
semimetals~\bibnote{In type-I Weyl semimetals, the major
contribution to the BCDM comes from the WPs,
whereas the rest of the Fermi surface makes a negligible
contribution due to the isotropic group velocities
near the WPs. Therefore, we notice a considerable change
in the BCDM with respect to the chemical potential in T$_d$-MoTe$_2$~\cite{Facio2018, Binghai2018}.}.
Fig.~\ref{fig:BCDM}(c) shows that $D_{yx}$ is peaked near the Fermi level,
while $D_{xy}$ exhibits oscillating behavior.
At $E_{F}$, $D_{xy}$ and $D_{yx}$ are estimated
to be 0.04 and 0.17, respectively~\bibnote{Note that the BCDM is a dimensionless quantity in three dimensions.}.
These values are relatively smaller than the corresponding
$D_{xy}=0.8$ and $D_{yx}=-0.7$ reported for T$_d$-MoTe$_2$
by Zhang et al.~\cite{Binghai2018}.
The main reason behind this difference is the strong
sensitivity of the Fermi surface to the on-site Hubbard $U$
of Mo $4d$ electrons~\cite{NXu_PRL2018, Aryal_PRB2019}, which
was not taken into account in the previous study~
\bibnote{Only four WPs are reported in Ref.~\cite{Binghai2018},
in contrast to the twelve (4 W1 and 8 W2) obtained in our case.}.

From Eq.~\ref{eq:chi}, the nonlinear conductivity tensor has
nonzero terms
$\chi_{xxz}=-\chi_{zxx}$ associated with $D_{xy}$, and
$\chi_{zyy}=-\chi_{yyz}$ associated with $D_{yx}$.
In view of the significant peak in $D_{yx}$ near $E_{\rm F},$
one interesting measurement would be the observation
of a transverse current $j_z$ induced by an
oscillating electric field along $y$ direction.
In the $\omega\rightarrow0$ limit,
an external electric field applied along
the $y$ direction, i.e., the chain direction, generates an
out-of-plane current $j_z^0\,=\,2\chi_{zyy}|\mathcal{E}_y|^2$.
If one can raise the electron chemical potential {\it via} gating,
the transverse current $j_z^0$ is predicted to rapidly reach its
maximum and then decrease, and eventually reverse its sign.

Here, we stress that a structural transition from the
T$_d$-A to T$_d$-B phase flips the sign of $D_{ab}$ while keeping its
magnitude intact, thus, allowing one to distinguish
between the two variants of polar T$_d$ phases~\bibnote{We do not
notice significant differences in the magnitudes of $D_{ab}$ for the
intermediate structures along the polarity reversal path, although new pairs
of WPs get created/annihilated as a function of $\lambda$.
This is due to the fact that most of the newly created WPs 
have relatively smaller tilt of Weyl cone
compared to W1, thus yielding minimal changes to the
overall BCDM.}.
For this purpose, observation of $D_{xy}$ via
$j_z^0\,=\,2\chi_{zxx}|\mathcal{E}_x|^2$ may be most
suitable, since the sign of $D_{xy}$ is less sensitive to the electron
chemical potential.

An interesting aspect of the nonlinear Hall conductivity in this
system is that, because the surfaces have lower symmetry than
the bulk, new components of the $D$ tensor are activated
at the surface.  In particular,
the glide mirror $\mathcal{M}_x\,\mathcal{T}({\vec{c}/2})$
is broken at the (001) cleavage surface.
Recall that the $D_{yz}$ and $D_{zy}$ tensor elements were argued
to vanish in the bulk because of this glide mirror, but they need
not vanish at the surface. Thus, response currents associated
with the conductivity tensor elements $\chi_{yyx}\,=\,-\chi_{xyy}$
and $\chi_{xzz}\,=\,-\chi_{zzx}$ are allowed.  While we can
confidently predict the existence of such currents, we are not
currently in a position to compute the surface $D$ tensors
quantitatively.  This observation thus provides a
challenge for future efforts at both theoretical prediction and
experimental detection of surface nonlinear Hall responses.

We may also consider the symmetries that remain in
the exfoliated few-layer limit.
In fact, the $\chi_{yyx}=-\chi_{xyy}\propto D_{yz}$
tensor elements are the only ones to survive in this limit.
The other terms, proportional to $\Omega_{x}$ or $\Omega_{y}$,
are not well defined in two dimensions. Therefore, measuring the in-plane
nonlinear Hall conductivity of MoTe$_2$ with respect to
the film thickness may reveal a noticeable transition
from the film to the surface responses.

In principle, one can utilize the nonlinear response current generated
due to the rapid fluctuation of $D_{yx}$ and its sign reversal
near the Fermi level as a function of the chemical
potential to devise a nonlinear Hall transistor for practical applications.
Moreover, recent experiments~\cite{MYZhang_PRX2019, EdbertSie_wte2Nat2019}
demonstrated an ultrafast optical control over T$_d$ and 1T$'$ structural phase
transitions, hence, an ultrafast topological optical switch can be designed using
the nonlinear quantum Hall property of MoTe$_2$, where T$_d$ (1T$'$) phase
can act as an {\sc ON} ({\sc OFF}) state.

Unlike in polar insulators, in which the switching of polarity
is immediately manifested in a polarization switching current, a
corresponding experimental response is missing in the case of
polar metals. Here, we propose that the nonlinear Hall effect
may serve as a potential experimental response to detect the polarization
direction/switching in polar metals, particularly, in nonmagnetic Weyl semimetals.
As demonstrated above, the polarization switching in
Weyl semimetals is always accompanied by the reversal of
the nonlinear Hall response.

In summary, we explain the intricate structural phase transitions
in MoTe$_2$ by defining a high-symmetry nonpolar phase T$_0$
that exhibits a higher-order topology. We unveil the connection
between the Weyl phase and the higher-order topological phase
in MoTe$_2$. We report that WPs can be readily created/annihilated,
manipulated, and switched by controlling the structural
phase transitions between the two polar variants of
the T$_d$ phase. Since this structural phase
switching has already been experimentally achieved,
and is shown to be reversible~\cite{huang2019polar},
MoTe$_2$ offers a promising platform
to harness the dynamics of Weyl fermions for technological applications.
We also report on the presence of a
tunable nonlinear Hall effect in T$_d$-MoTe$_2$,
and discuss the potential applications of this effect in
designing ultrafast topological optical switches and transistors.
Lastly, we propose that the nonlinear Hall effect can be
utilized as a potential experimental response to detect polarization
direction/switching in polar metals or semimetals that inherit large
 concentrations of Berry curvature near the
Fermi energy, e.g, in nonmagnetic Weyl semimetals.

\begin{acknowledgments}
We thank Fei-ting Huang and
Sang-Wook Cheong for fruitful discussions.
This work was supported by ONR Grants N00014-16-1-2951 and N00014-19-1-2073.
\end{acknowledgments}

{\it Note:} Supplemental Materials can be obtained from the corresponding author on reasonable request.

\bibliography{bibmote2.bib}

\end{document}